\documentclass[conference]{IEEEtran}
\IEEEoverridecommandlockouts
\usepackage{cite}
\usepackage{amsmath,amssymb,amsfonts}
\usepackage{algorithmicx}
\usepackage{algorithm}
\usepackage{graphicx}
\usepackage{epstopdf}
\usepackage{textcomp}
\usepackage{xcolor}
\def\BibTeX{{\rm B\kern-.05em{\sc i\kern-.025em b}\kern-.08em
    T\kern-.1667em\lower.7ex\hbox{E}\kern-.125emX}}
\begin{document}

\title{Max-Min Energy-Efficiency Fair Optimization in STAR-RIS Assisted System
}

\author
{\IEEEauthorblockN{1\textsuperscript{st} Ruitianyi~Lu}
\IEEEauthorblockA{\textit{School of Information Engineering} \\
\textit{Wuhan University of Technology}\\Wuhan, China \\
lrty@whut.edu.cn}
\and
\IEEEauthorblockN{2\textsuperscript{nd} Kehao~Wang}
	\IEEEauthorblockA{\textit{School of Information Engineering} \\
		\textit{Wuhan University of Technology}\\Wuhan, China \\
		wkh@whut.edu.cn}
}

\maketitle

\begin{abstract}
In this paper, we investigate simultaneously transmitting and reflecting reconfigurable intelligent surfaces (STAR-RIS), which enables to communicate with users both sides by transmitting signals to users forward and reflecting signals to users backward simultaneously. We consider a communication system with a STAR-RIS, a base station (BS) and many users to maximize the minimum user energy efficiency (EE) by jointly optimizing the active beamforming, transmission and reflection coefficients with BS power consumption limited. To solve this optimization problem efficiently, we divide it into two subproblems to optimize the transmitting beamforming matrix and phase shifts of STAR-RIS seperately. With the two subproblems fixed, an Alternating Optimization (AO) method is proposed to solve the maximize minmum user EE fair optimization problem. Numerical results can demonstrate that STAR-RIS behaves better than traditional reflecting-only RIS, and the algorithm we desigend can maximize the minum EE problem efficienty to ensure user fairness.
\end{abstract}


\section{Introduction}
  The upcoming 6G wireless network demands high spectral efficiency (SE) and massive connectivity. Since high channel capacity has been considered widely, energy efficiency (EE), defined as the ratio of SE over power consumption, becomes the focus of our research. Traditional relay in wireless communication network inevitably brings a lot of hardware loss, thus reconfigurable intelligent surface (RIS) is considered a promising new technology for reconfiguring the wireless propagation environment through software controlled reflection.  
 
Reflecting-only RISs have been fully investigated in
  wireless communication systems to enhance transmission quality \cite{2018Intelligent} and maximize RIS-assited system EE\cite{2020Energy,2019Reconfigurable}. However, RISs who can only reflect incident signals face significant spatial limitations that all the users have to be the same side of RIS. In order to overcome this absolute space limitation and to meet the needs of our 360-degree users, we propose the concept of simultaneous transmitting and reflecting RIS (STAR-RIS) in this letter. While retaining the reflected signal characteristics of RIS, STAR-RIS can also transmit incident signals to users behind itself at the same time. Based on its own electromagnetic physical characteristics, both the reflected and transmitted signals can be adjusted and reconstructed by reflection and transmission coefficients of STAR-RIS. The users in front of STAR-RIS receive reflected signals, while users behind it receiving transmitted signals, thus enabling 360-degree communication.  

As an arising technology, there are plenty of papers introducting STAR-RIS assisted communication systems. A general hardware model for STAR-RISs is presented in \cite{2021STAR}. Channel models are proposed for the
near-field and the far-field scenarios, base on which the diversity
gain of the STAR-RIS is analyzed and compared with that of the conventional reflecting-only RIS. Numerical simulations can verify analytical results and to demonstrate that full diversity order can be achieved on both sides of the STAR-RIS. Based on its unique simultaneous transmission reflection mechanism, three practical transmission protocols are shown in \cite{2021Simultaneously}, namely energy splitting (ES), mode switching (MS), and time switching (TS). Literature \cite{2021Simultaneously} desicribe a STAR-RIS aided downlink communication system considered for both unicast and multicast transmission, where a multi-antenna base station (BS) sends information to two users on each side of the STAR-RIS. A power consumption minimization problem for the joint optimization of the active beamforming at BS and the passive transmission and reflection beamforming at the STAR-RIS is formulated for each of the proposed operating protocols, subject to communication rate constraints of the users.
Numerical results reveal that the required power consumption for both scenarios is significantly reduced by employing the proposed STAR-RIS
instead of conventional reflecting/transmiting-only RIS. Paper \cite{2021Coverage} aims to characterize the fundamental coverage range of STAR-RIS aided communication networks. An AP communicates with one transmitted user and one reflected user employing both non-orthogonal multiple access (NOMA) and orthogonal multiple access (OMA) with the aid of an STAR-RIS \cite{2021Coverage}. A sum coverage range maximization problem is formulated for each multiple access scheme. 

In this paper, we propose a maximize minmum user EE optimization to ensure communication fair among transmitting and reflecting users \cite{Li2013Transmit}. According to user's relative location, they are divided to reflecting users and transmitting users depending on whether it's in front or behind STAR-RIS. Given BS limited power consumption, we jointly optimize the power allocating matrix, transmitting and reflecting coeffients matrixs of STAR-RIS to maximize the minal EE problem. By decomposing it into two subproblems, the semi-definite
relaxation (SDR) approach \cite{Luo2010Semidefinite,2002Smoothing}and  successive convex approximation (SCA) \cite{2014Parallel} is applied to optimize the beamforming vectors at the BS and the phase shifts of STAR-RIS. Finally, AO algorithm is utilized to solve the original max-min EE optimization problem.

The layout of this work is organized as follows: STAR-RIS and relative works are introduced in Section \uppercase\expandafter{\romannumeral1}. Section \uppercase\expandafter{\romannumeral2} investigates the signal model and power consumption model of STAR-RIS aissited communication system. An efficient optimization algorithm is proposed for the energy splitting (ES) scheme in Section \uppercase\expandafter{\romannumeral3} to solve the max-min EE optimization problem. Section \uppercase\expandafter{\romannumeral4} shows the numerical results.

\textit{Notation}: $a$ is a scalar, $\mathbf{a}$ is a vector, and $\mathbf{A}$  is a matrix. $\mathbf{A}^T$, $\mathbf{A}^H$, $\mathbf{A}^{-1}$, and $\|\mathbf{A}\|_F$ denote transpose, Hermitian (conjugate transpose), inverse, and Frobenius norm of $ \mathbf{A} $, respectively. $\mathrm{Re}\{\cdot\}$, $|\cdot|$, $(\cdot)^*$ and $\mathrm{arg}(\cdot)$ denote the real part, modulus, conjugate and the angle of a complex number, respectively. $\text{tr}(\cdot)$ denotes the trace of a matrix and $\mathbf{I}_n$ (with $n\geq2$) is the $n\times n$ identity matrix. $ \mathrm{diag}(\mathbf{a})$ is a diagonal matrix with the entries of $\mathbf{a}$ on its main diagonal. $\mathbf{A}\succeq\mathbf{B} $ means that $\mathbf{A}-\mathbf{B}$ is positive semidefinite. $j\triangleq\sqrt{-1}$ is the imaginary unit.
\section{System Model}
\subsection{System Model}
\begin{figure}[htbp]\vspace{-1mm}
	\centering
	\includegraphics[width=80mm]{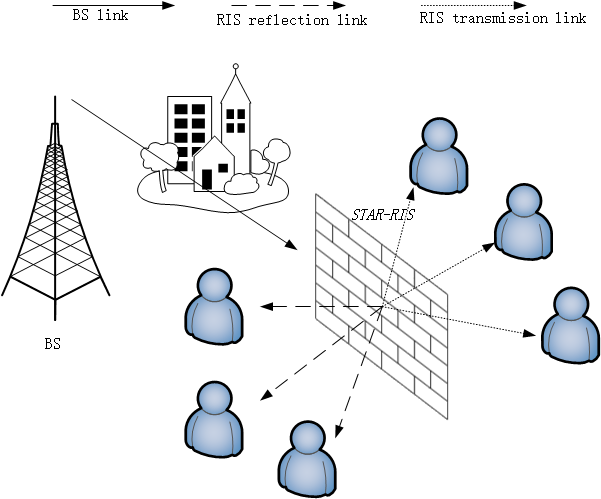}  \vspace{-3mm}
	\caption{The considered RIS-based multi-user MIMO system comprising of a $M$-antenna BS simultaneously serving in the downlink $K$ single-antenna users.}
	\label{fig:Estimation_Scheme} \vspace{-3mm}
\end{figure}

Consider a downlink communication system consisting of one BS equipped with M antenna element and K single-antenna mobile users. In the STAR-RIS aided system, the BS broadcasts independent superposed data streams to
the $k$ users with beamforming vectors \{$v_1$, $v_2$,..., $v_k$\}, where $v_k$ is the active beamforming vector for user $k$. We assume that this communication takes place via RIS with N reflecting elements simultaneously transmitting and reflecting signals emmitted from BS to users on both sides, as illustrated in Fig. 1. The direct signal path between BS and mobile users is intercepted due to the buildings between them.

The received signal at the $k$-th user is given by
\begin{equation}
\begin{aligned}
y_k = g_{k} \Phi H v_{k} s_k + \sum^K_{\bar{k} = 1,\neq k}  ( g_{k} \Phi H v_{\bar{k}} ) + \sigma^2
\end{aligned}
\end{equation}
where $g_{k} \in  \mathbb{C}^{1 \times N}$ suggests the receiving matrix between RIS and $k$-th user, $H \in \mathbb{C}^{N \times M} $ represents the transmission characteristics from the BS to the RIS. $v_{k}\in \mathbb{C}^{M \times 1} $ represent the transmit beamforming vector from BS to $k$ user, $\Phi\in \mathbb{C}^{N \times N} $ is the RIS coefficient diagonal matrix which is defined as
$\Phi_k=\begin{cases}
\Phi_t,\; \text{if user $k$ is located in the transimission area}\\
\Phi_r,\; \text{if user $k$ is located in the reflection area}\end{cases}$

Spectrum efficiency
\begin{equation}
\begin{aligned}\label{r1}
R_k = \log_2\left(1+
\frac{| g_{k} \Phi Hv_{k}|^2 }
{\sum^K_{\bar{k}=1,\neq k} | g_{k} \Phi Hv_{\bar{k}}|^2  + \sigma^2} \right)
\end{aligned}
\end{equation}

For convenience, let us define the equivalent channel spanning from the BS to the $k$-th user $h_{k} = g_{k} \Phi H$ .

\subsection{User Power Consumption Model}\label{sec:problem}
Composed of the BS transmit power, the hardware static power consumed in the BS, mobile user terminals and RIS, $k$-th user will consume the power $\mathcal{P}_{k}$ to connect with BS, which can be expressed as
\begin{align}\label{power_model1}
\mathcal{P}_k\triangleq \alpha p_k + P_{{\rm h},k}+ P_{\rm B} + P_{\rm R},
\end{align}
where $\alpha \triangleq \nu^{-1}$, $\nu$ is the efficiency of the transmit power amplifier, $p_k$ represents the transmit power,  $P_{{\rm h},k}$ denotes the hardware static power of $k$-th UE, $P_{\rm B}$ and $P_{\rm R}$ denote the total hardware static power consumption by BS and RIS, respectively. Furthermore, the total power consumption of each user can be expressed as
\begin{align}\label{power_model2}
\mathcal{P}_{\rm k}= ||v_{k}||^2_F + P
\end{align}
where $\mathcal{P}_{\rm k}$ is devided into transmitting energy consumption and static circuit power consumption $P$ for reciever $k$.
Define the EE for each user as the data rate for user k per total dissipation power which is given by
\begin{align}\label{power_model3}
\mathcal{\eta}_{\rm k}(v_k,\Phi)= \frac{R_k(v_k,\Phi)}{\mathcal{P}_k(v_k)}
\end{align}
\subsection{Design Problem Formulation} \label{sec:problem}
We are interested in maximizing the minimum EE of
users to guarantee fairness among them by jointly optimizing
the transmit beamforming vectors at the BS, and the transmitting and reflectionphase shifts at the RIS, appearing in the diagonal of $\mathbf{\Phi}=\mathrm{diag}[\phi_1,\phi_2,\ldots,\phi_N]$. The BS release constant transmit power $P_{max}$ continuously.

The considered  maximize  power problem is expressed as follows:
\begin{subequations}\label{p1}
	\begin{align}
	\mathcal{P}_1: &\displaystyle \max_{\mathbf{\Phi},v_{k}}\min_{k} \frac{R_k}{||v_{k}||^2_F + P} \label{p1a}\\
	\text{s.t.}
	&\;\quad\;\; \sum_{k=1}^K  ||v_{k}||^2_F \leq P_{max} \label{p1b}\\
	&\;  \quad\;\;R_k \geq \mathbf{R},\; \forall k\label{p1c}\\
	&\;\quad\;\; [\Phi]_n = \sqrt{\beta_n^k}e^{j\theta_n^k}, \beta_n^t + \beta_n^r = 1,\notag\\&\theta_n^t,\theta_n^r \in[0,2\pi), \;\forall n=1,2,\ldots,N, \label{p1d}
	\end{align}
\end{subequations}
where $R$ denotes the individual QoS constraint of the user, $P$ represents the total static power in the system. Constraint($\ref{p1b}$) nsures that the sum of the powers to reach the RIS is equal to a fixed value. Constraint($\ref{p1c}$) promises each user can get qualified signal whose SNR reached $\mathbf{R}$. Constraint($\ref{p1d}$) explains physical constraints to the transmission and reflection
coefficients of every STAR-RIS elements.

\section{Optimization Formulation And Analysis}
In this section, we investigate Alternating Optimization(AO) approach to handle problem $\mathcal{P}_1$ efficiently by dividing it into two subproblems.
\subsection{ Transmit Beamforming Matrix Optimization with Fixed Phase Shift Matrix}\label{18MM}
First, we apply non-liner (NL) fractional programming theory \cite{1977Fractional} to solve the fractional problem. Hence, $\mathcal{P}_1$ can be formulated as:
\begin{subequations}\label{12bb}
	\begin{align}
	\mathcal{P}_1: &\displaystyle \max_{\mathbf{\Phi},v_{k}}\min_{k}  \lambda(v_k,\Phi) = \frac{R_k(v_k,\Phi)}{\mathcal{P}_k(v_k)} \label{1a}\\
	\text{s.t.}
	&\;\quad\;\; \sum_{k=1}^K  ||v_{k}||^2_F \leq P_{max} \label{Prob:bResAllpower}\\
	&\;  \quad\;\;R_k \geq \mathbf{R},\; \forall k\label{Prob:cResAllpower}\\
	&\;\quad\;\; [\Phi_c]_{n,n}=\beta_n^c,\beta_n^t+\beta_n^r =1,\notag\\ &|\Phi_c|=1,\;\forall n=1,2,\ldots,N,c\in\{t,r\} \label{Prob:dResAllpower}
	\end{align}
\end{subequations}
According to Dinkelbach algorithm \cite{2016Max}, the fractional problem has
the following form:
\begin{align}
\lambda^* = \max_{\mathbf{\Phi},v_{k}}\min_k\frac{R_k(v_k,\Phi)}{\mathcal{P}_k(v_k)}
\end{align} 
Theorem 1: The optimal solution to the problem in ($\ref{12bb}$) is achieved if and only if:
\begin{align}\label{1n}
\max_{\mathbf{\Phi},v_{k}}\min_{k}\{R_k(v_k,\Phi) - \lambda\mathcal{P}_k(v_k)\}\notag\\=\min_{k}\{R_k(v_k,\Phi) - \lambda\mathcal{P}_k(v_k)\}=0
\end{align}
It is not hard to find that equivalent problem ($\ref{1n}$) is strictly decreasing in $\lambda$. To facilitate the solution design and to smooth the objective function, an auxiliary variable $\mu$ is introduced so that
the equivalent form ($\ref{p1a}$) can be expressed as:
\begin{subequations}\label{Prob:ResAllpower}
	\begin{align}
	\mathcal{P}_2: &\displaystyle \max_{\mathbf{\Phi},v_{k},\lambda,\mu}\mu\label{1a}\\
	\text{s.t.}
	&\;\quad\;\; R_k(v_k,\Phi) - \lambda\mathcal{P}_k(v_k)\geq \mu\label{Prob:bResAllpower}\\
	&\;\quad\;\;\sum_{k=1}^K  ||v_{k}||^2_F \leq P_{max} \label{Prob:cResAllpower}\\
	&\;  \quad\;\;R_k \geq \mathbf{R},\; \forall k\label{Prob:dResAllpower}\\
	&\;\quad\;\; [\Phi_c]_{n,n}=\beta_n^c,\beta_n^t+\beta_n^r =1,\notag\\ &|\Phi_c|=1,\;\forall n=1,2,\ldots,N,c\in\{t,r\} \label{Prob:eResAllpower}
	\end{align}
\end{subequations}
Up to now we have finished the hard fractional objective function. By applying the AO algorithm with fixed phase shifts, problem ($\mathcal{P}_2$) can be reformulated
as the following equivalent problem:
\begin{subequations}\label{Prob:ResAllpower}
	\begin{align}
	&\displaystyle \max_{v_{k},\lambda,\mu}\mu\label{1a}\\
	\text{s.t.}
	&\;\quad\;\; R_k(v_k,\Phi) - \lambda\mathcal{P}_k(v_k)\geq \mu\label{Prob:bResAllpower}\\
	&\;\quad\;\;\sum_{k=1}^K  ||v_{k}||^2_F \leq P_{max} \label{Prob:cResAllpower}\\
	&\;  \quad\;\;R_k \geq \mathbf{R},\; \forall k\label{Prob:dResAllpower}
	\end{align}
\end{subequations}
To optimize this sub-problem, the semi-definite relaxation (SDR) approach is applied to optimize the beamforming vectors at the BS. Given the reflection and transmitting matrix $\mathbf{\Phi}$ of RIS , $h_{k} = g_{k} \mathbf{\Phi}\mathbf{H}$ is certain therefore. Define $\mathbf{V}_k = v_kv_k^H,\mathbf{H}_k= h_kh_k^H,\forall k$, where matrix $\mathbf{V}_k$ is semi-definite and satisfies rank($\mathbf{V}_k$) $\leq$ 1. 
By applying SDR technique \cite{Luo2010Semidefinite}, we have:
\begin{align}
R_k(v_k,\Phi) &= log_2(\text{Tr}(\mathbf{H}_k\mathbf{V}_k) + \sum^K_{\bar{k}=1,\neq k} \text{Tr}(\mathbf{H}_{k}\mathbf{V}_{\bar{k}}) + \sigma^2) \notag\\&- log_2(\sum^K_{\bar{k}=1,\neq k} \text{Tr}(\mathbf{H}_{k}\mathbf{V}_{\bar{k}}) + \sigma^2),\forall k
\end{align}
\begin{align}
\mathcal{P}_k(v_k) = \text{Tr}(\mathbf{V}_k) + P
\end{align}
Thus, the subproblem can be restated as:
\begin{subequations}\label{p3}
	\begin{align}
	\mathcal{P}_3:&\displaystyle \quad\;\;\max_{\mathbf{V}_k,\mu}\mu \\
	\text{s.t.}
	&\;\quad\;\; log_2(\text{Tr}(\mathbf{H}_k\mathbf{V}_k) + \sum^K_{\bar{k}=1,\neq k} \text{Tr}(\mathbf{H}_{k}\mathbf{V}_{\bar{k}}) + \sigma^2) \notag\\&- log_2(\sum^K_{\bar{k}=1,\neq k} \text{Tr}(\mathbf{H}_{k}\mathbf{V}_{\bar{k}}) + \sigma^2) - \lambda(\text{Tr}(\mathbf{V}_k) + P)\geq \mu\notag\\
	&\;\quad\;\;\sum_{k=1}^K  \text{Tr}(\mathbf{V}_k)\leq P_{max} \label{Prob:cResAllpower}\\
	&\;  \quad\;\;\frac{\text{Tr}(\mathbf{H}_k\mathbf{V}_k)}{2^\mathbf{R}-1} - \sum^K_{\bar{k}=1,\neq k} \text{Tr}(\mathbf{H}_{k}\mathbf{V}_{\bar{k}}) \geq \sigma^2,\; \forall k\label{Prob:dResAllpower}\\
	&\;  \quad\;\;\mathbf{V}_k \geq 0
	\end{align}
\end{subequations}
The proposed algorithm for solving (P3) is given in Algorithm 1, to settle this sub-problem.
\begin{algorithm}[!t]
	\caption{Dinkelbach Algorithm for $\mathbf{P_3}$.}
	\begin{algorithmic}[4]
		\State \textbf{Initialization:} iterative number n = 1, maximum number of iterations $n_{max}$, accuracy $\epsilon$. Setting the maximum energy efficiency $\lambda^{(1)}$=0.
		\State \textbf{REPEAT}
		\State Solve problem (P6) with $\lambda^{(n)}$ and obtain $\{v_k^{(n)},\mu^{(n)} \}$.
		\State \textbf{if} $|\min\limits_k R_k(v_k^{(n)}) - \lambda^{(n)}\mathcal{P}_k(v_k^{(n)})| \leq \epsilon$
		\State  \textbf{RETURN} $v_k^*,\mu^* = v_k^{(n)},\mu^{(n)}$, $\lambda^* = \min\limits_k\frac{R_k(v_k^{(n)})}{\mathcal{P}_k(v_k^{(n)})}$.
		\State \textbf{Else}  Update $\lambda^{(n+1)}=\min\limits_k\frac{R_k(v_k^{(n)})}{\mathcal{P}_k(v_k^{(n)})}$.
		\State \textbf{UNTIL} $n \geq n_{max}$.
	\end{algorithmic}
\end{algorithm}
With $\lambda$ solved, we notice that the relaxed $\mathcal{P}_3$ is a standard semi-definite programming (SDP) \cite{2007A} which can be solved by using CVX \cite{2008CVX}. Hence, a tractable form of (P2) is obtained where we
handle (P3) instead of (P2). 

Up to now, we have figured out the beamforming vectors at the BS. Next, we attempt to optimize the reflection and transmitting matrix at the STAR-RIS by utilizing the AO algorithm \cite{Li2013Transmit}.

\subsection{Transmitting and Reflection Phase Shifts Matrix Optimization with Fixed Transmit Beamforming}\label{20MM}
Given the transmitting beamforming vetor $v_k$, we go back to our original problem model ($\ref{p1}$). Denote $b_k = diag(g_k)Hv_k$ and $\mathbf{\Phi}_c =\phi_c\phi_c^H$. where $\mathbf{\Phi}_c \geq 0$, rank($\mathbf{\Phi}_c$) =1 and [$\mathbf{\Phi}_c$]$_{n,n}$ = $\beta_n^c$, c $\in \{t,r\}$. All STAR-RIS elements simultaneously operate in transmitting and reflecting modes under the energy splitting (ES) protocol. Hence, we have
\begin{align}
| g_{k} \Phi Hv_{k}|^2 = |u_k^H b_k|^2 = \text{Tr}(\mathbf{U_k}\mathbf{B_k})
\end{align}
where $\mathbf{B_k}$ = $b_kb_k^H$ and the matrix $\mathbf{U_k}$ is defined as:
\begin{align*}
\mathbf{U_k} = \left\{
\begin{aligned}
&\mathbf{\Phi}_t,  \text{if user k is located in the transmitting space}\\
&\mathbf{\Phi}_r,  \text{if user k is located in the reflection space}
\end{aligned}
\right\}
\end{align*}
Thus, the transmission and reflection phase shifts optimization problem in ($\ref{p1}$) with fixed active beamforming vectors can be formulated as:
\begin{subequations}\label{p4}
	\begin{align}
	\mathcal{P}_4:&\displaystyle \quad\;\;\max_{\mathbf{\Phi}_c,\mu}\mu \\
	\text{s.t.}
	&\;\quad\;\; log_2(\text{Tr}(\mathbf{U}_k\mathbf{B}_k) + \sum^K_{\bar{k}=1,\neq k} \text{Tr}(\mathbf{U}_k\mathbf{B}_{\bar{k}}) + \sigma^2) \notag\\&- log_2(\sum^K_{\bar{k}=1,\neq k} \text{Tr}(\mathbf{U}_k\mathbf{B}_{\bar{k}}) + \sigma^2) - \lambda(\text{Tr}(\mathbf{V}_k) + P)\geq \mu\notag\\
	&\;  \quad\;\;\frac{\text{Tr}(\mathbf{U}_k\mathbf{B}_k)}{2^\mathbf{R}-1} - \sum^K_{\bar{k}=1,\neq k} \text{Tr}(\mathbf{U}_k\mathbf{B}_{\bar{k}}) \geq \sigma^2,\; \forall k\\
	&\;\quad\;\; \beta_n^t + \beta_n^r = 1\\
	&\;\quad\;\; [\mathbf{\Phi}_c]_{n,n} = \beta_n^c, \mathbf{\Phi}_c \geq 0\\
	&\;\quad\;\; rank(\mathbf{\Phi}_c) =1\label{p4e}
	\end{align}
\end{subequations}
where c $\in \{t,r\}$. It can be concluded that ($\ref{p4e}$) makes $\mathcal{P}_4$ non-convex. According to \cite{2020Exploiting,2020Joint}, the non-convex rank-one constraint ($\ref{p4e}$) can be replaced by the following relaxed convex constraint:
\begin{align*}
\xi_{max}(\Phi_c) \geq \varepsilon^{(\tau)}\text{Tr}(\Phi_c)
\end{align*}
where $\xi_{max}(\Phi_c)$ denotes the maximum eigenvalue of matrix $\Phi_c$, $\varepsilon^{(\tau)}$ is a relaxation parameter in the $\tau$-th iteration, which controls $\xi_{max}(\Phi_c)$ to trace ratio of $\Phi_c$. Specifically, $\varepsilon^{(\tau)}$  = 0 indicates that the rank-one constraint is dropped; $\varepsilon^{(\tau)}$ = 1 is equivalent to the rank-one constraint. Therefore, we can increase $\varepsilon^{(\tau)}$ from 0 to 1 sequentially via iterations to gradually approach a rank-one solution. It is noted that $\xi_{max}(\Phi_c)$ is not differentiable, i.e., non-smooth. The following approximation expression can be used to approximate $\xi_{max}(\Phi_c)$:
\begin{align*}
\xi_{max}(\Phi_c) = e_{max}^H(\Phi_c^{(\tau)})\Phi_ce_{max}(\Phi_c^{(\tau)})
\end{align*}
where $e_{max}^H(\Phi_c^{(\tau)})$ is the eigenvector corresponding to the maximum eigenvalue of $\Phi_c^{(\tau)}$.
Therefore, ($\ref{p4}$) is transformed to the following relaxed problem:
\begin{subequations}\label{p5}
	\begin{align}
	\mathcal{P}_5:&\displaystyle \quad\;\;\max_{\mathbf{\Phi}_c,\mu}\mu \\
	\text{s.t.}
	&\;\quad\;\; log_2(\text{Tr}(\mathbf{U}_k\mathbf{B}_k) + \sum^K_{\bar{k}=1,\neq k} \text{Tr}(\mathbf{U}_k\mathbf{B}_{\bar{k}}) + \sigma^2) \notag\\&- log_2(\sum^K_{\bar{k}=1,\neq k} \text{Tr}(\mathbf{U}_k\mathbf{B}_{\bar{k}}) + \sigma^2) - \lambda(\text{Tr}(\mathbf{V}_k) + P)\geq \mu\notag\\
	&\;  \quad\;\;\frac{\text{Tr}(\mathbf{U}_k\mathbf{B}_k)}{2^\mathbf{R}-1} - \sum^K_{\bar{k}=1,\neq k} \text{Tr}  (\mathbf{U}_k\mathbf{B}_{\bar{k}}) \geq \sigma^2,\; \forall k\\
	&\;\quad\;\; \beta_n^t + \beta_n^r = 1\\
	&\;\quad\;\; [\mathbf{\Phi}_c]_{n,n} = \beta_n^c, \mathbf{\Phi}_c \geq 0\\
	&\;\quad\;\; e_{max}^H(\Phi_c^{(\tau)})\Phi_ce_{max}(\Phi_c^{(\tau)}) \geq \varepsilon^{(\tau)}Tr(\Phi_c)\label{p5e}
	\end{align}
\end{subequations}
where $c \in \{t,r\}$. Problem ($\ref{p5}$) is a standard convex SDP, which can be solved efficiently by numerical solvers such as the SDP solver in CVX tool \cite{2008CVX}. The parameter $\varepsilon^{(\tau)}$ can be updated via \cite{2020Joint}
\begin{align}\label{CHA1}
\varepsilon^{(\tau+1)} = \min (1,\frac{\xi_{max}(\Phi_c^{(\tau)})}{Tr(\Phi_c^{(\tau)})} + \Delta^{(\tau)})
\end{align}
where $\Delta^{(\tau)}$ is the step size.
Details of the proposed sequential constaint relaxation method is presented in Algorithm 2.
\begin{algorithm}[!t]
	\caption{Sequential constraint relaxation for obtaining $\Phi_c$.}
	\begin{algorithmic}[4] 
		\State \textbf{Initialization:} $\Phi_c^{(0)}$, step size $\Delta^{(0)}$,error tolerance $\delta$, set relaxation parameter $\varepsilon^{(0)}$ = 0 and iteration index $\tau$= 0.
		\State \textbf{REPEAT}
		\State Solve problem ($\ref{p5}$) to obtain $\Phi_c$.
		\State \textbf{if} problem ($\ref{p5}$) is solvable:\\
		      \;\; Update $\Phi_c^{(\tau+1)}$ = $\Phi_c$.\\
		      \;\; Update $\Delta^{(\tau+1)}$ = $\Delta^{(0)}$.
		\State  \textbf{Else} \\
		\;\; Update $\Phi_c^{(\tau+1)}$ = $\Phi_c^{(\tau)}$.\\
		\;\; Update $\Delta^{(\tau+1)}$ = $\Delta^{(\tau)}/2$.
		\State \textbf{End} \\
		 \;\; Update $\tau$ = $\tau$ + 1.\\
		 \;\; Update $\varepsilon^{(\tau+1)}$ via ($\ref{CHA1}$).
		\State \textbf{UNTIL} $|1 - \varepsilon^{(\tau)}| \leq \delta$ and the objective value of problem ($\ref{p5}$) converge.
	\end{algorithmic}
\end{algorithm}
Based on the above discussions, we provide details of the proposed AO algorithm to solve the original max-min EE problem ($\ref{p1}$) in Algorithm 3.
\begin{algorithm}[!t]
	\caption{Alternating Optimization for problem ($\ref{p1}$).}
	\begin{algorithmic}[4] 
		\State \textbf{Initialization:} $v_k$, $\Phi_c$, error tolerance $\epsilon$, set iteration index i = 1.
		\State \textbf{REPEAT}
		\State Update active beamforming vectors $v_k$ via Algorithm 1.\\ Update passive beamforming vectors $\Phi_c$ via Algorithm 2.
		\State \textbf{UNTIL} the objective value of problem ($\ref{p1}$) converges.
	\end{algorithmic}
\end{algorithm}

\section{Numerical Results}
The environmental parameters are given as follows: the total number of users K = 6, BS antenna number M = 4, noise = -20dBm, SNR ratio threshold R = 6.6, RIS elements N = 30, system static power P = 5dBm and power emitted form BS Pmax = 50dBm. Having set the position parameters of the communication system, we verify the validity of our proposed approach through different user distributions. Via different reflected and transmitted users, we demonstrate that maximizing minimum user EE really works and ensure fariness between different type of users. Meanwhile, we also compare with traditional reflecting-only and transmitting-only RIS to see whether STAR-RIS has improved system performance indeed.

Firstly, we found that system EE grows with max energy power budget emitted by the BS in Fig 2, which means when Pmax is not big enough there's plenty room in the system channel. Fig 3 indicates that average Spectral efficiency (SE) increases with Pmax almost synchronically. However when Pmax retains a growing tunrning point, EE comes to a slow growth phase gradually reaching the maximum limit of the system. We can conclude that different numbers of transmitting and reflection users don't affect the EE optimization of the smallest user, and STAR-RIS performs better than reflecting-only and transmitting-only RIS.

\begin{figure}[!t] \vspace{-6mm}
	\begin{minipage}{9cm}
		\centering
		\includegraphics[width=8cm]{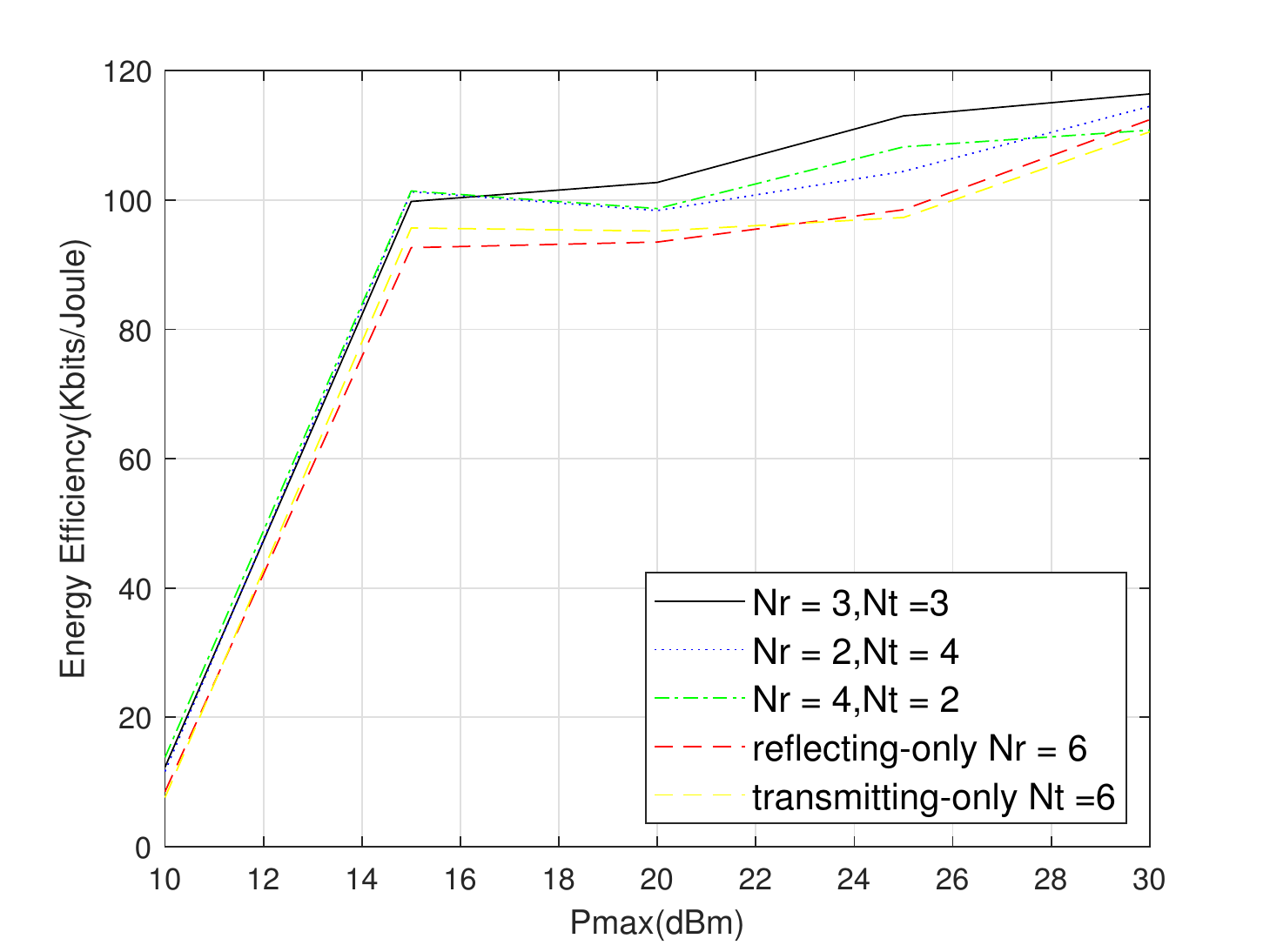}\vspace{-2mm}
		\caption{The minum user EE versus Pmax.}
		\label{fig:Estimation_Scheme} \vspace{-0.5mm}
	\end{minipage}
	\begin{minipage}{9cm}
		\centering
		\includegraphics[width=8cm]{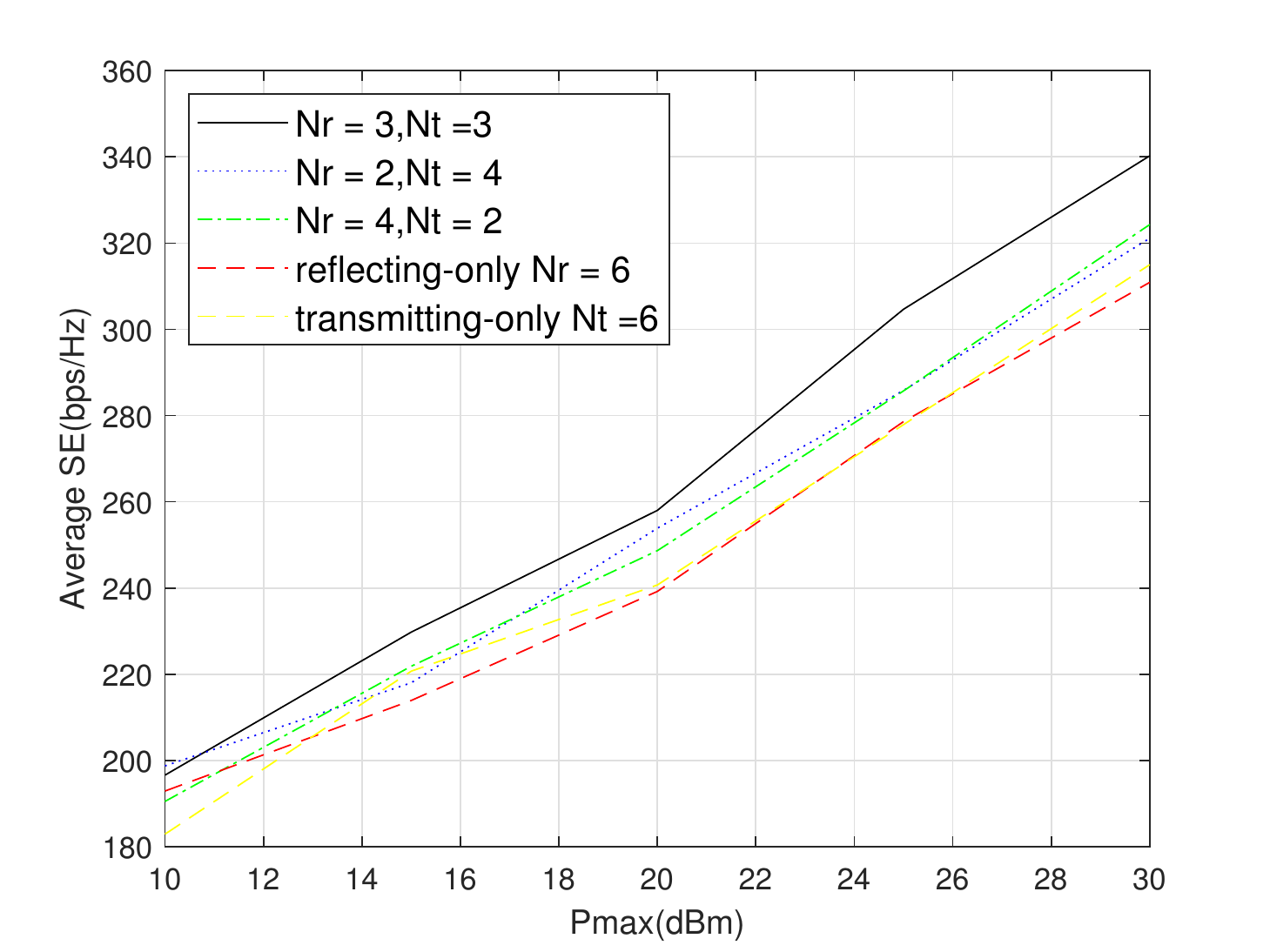}\vspace{-2mm}
		\caption{The average SNR versus Pmax.}
		\label{fig:Estimation_Scheme}\vspace{-1.5mm}
	\end{minipage}
\end{figure}

\begin{figure}[!t] \vspace{-6mm}
	\begin{minipage}{9cm}
		\centering
		\includegraphics[width=8cm]{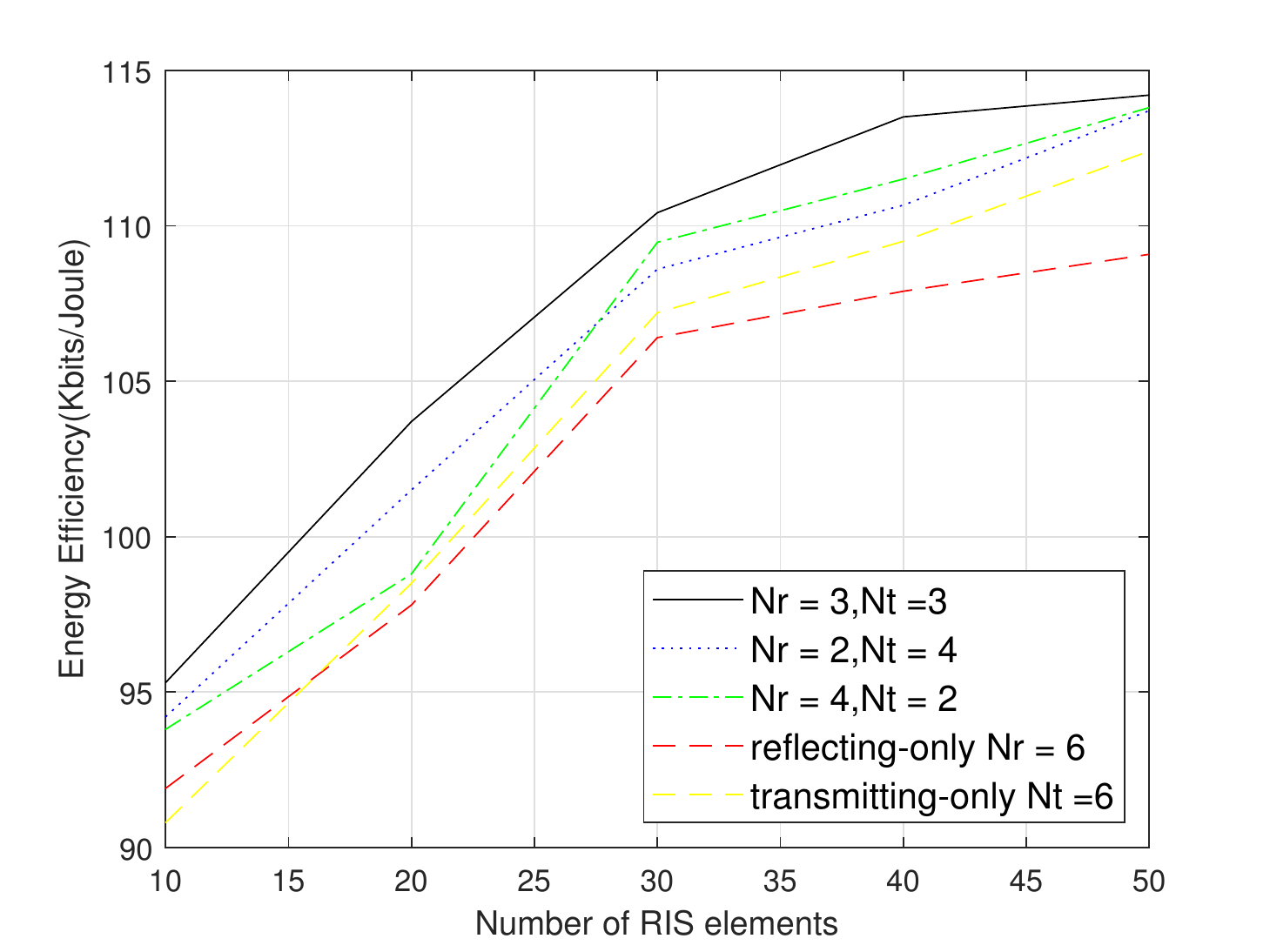}\vspace{-2mm}
		\caption{The minum user EE versus RIS elements number.}
		\label{fig:Estimation_Scheme} \vspace{-0.5mm}
	\end{minipage}
	\begin{minipage}{9cm}
		\centering
		\includegraphics[width=8cm]{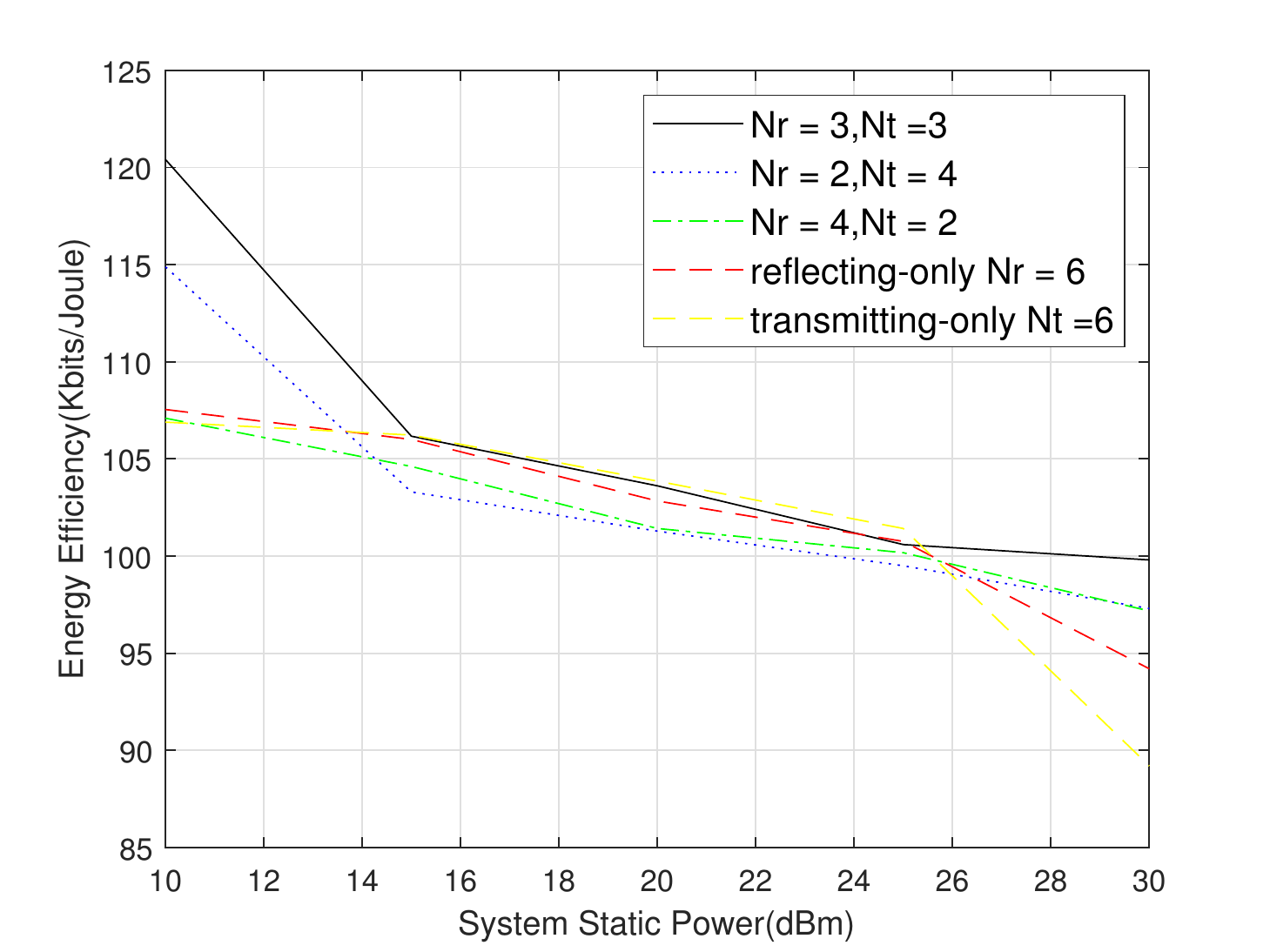}\vspace{-2mm}
		\caption{The average SNR versus system static power P.}
		\label{fig:Estimation_Scheme}\vspace{-1.5mm}
	\end{minipage}
\end{figure}
Fig 4 shows that EE is improved due to the increase of RIS element numbers. In the given uniform other conditions, STAR-RIS always perform better than reflecting-only and transmitting-only RIS. The more number of RIS pieces, the larger RIS area will be. So it can reflect/transmit more transmitting signals and reduce leakage. Thus improve the minal user EE. The increasing curves grow fast or slow because area is to the second power not linear and signals radiated from BS is uneven.

In addition to the positive trend, performance of different situations in the face of increased system static power is also documented in Fig 5. Simply rising hardware power consumption will no doubtly decreases user EE. It is worth mentioning that STAR-RIS declines more moderately.
\begin{figure}[!t] \vspace{-6mm}
	\begin{minipage}{9cm}
		\centering
		\includegraphics[width=8cm]{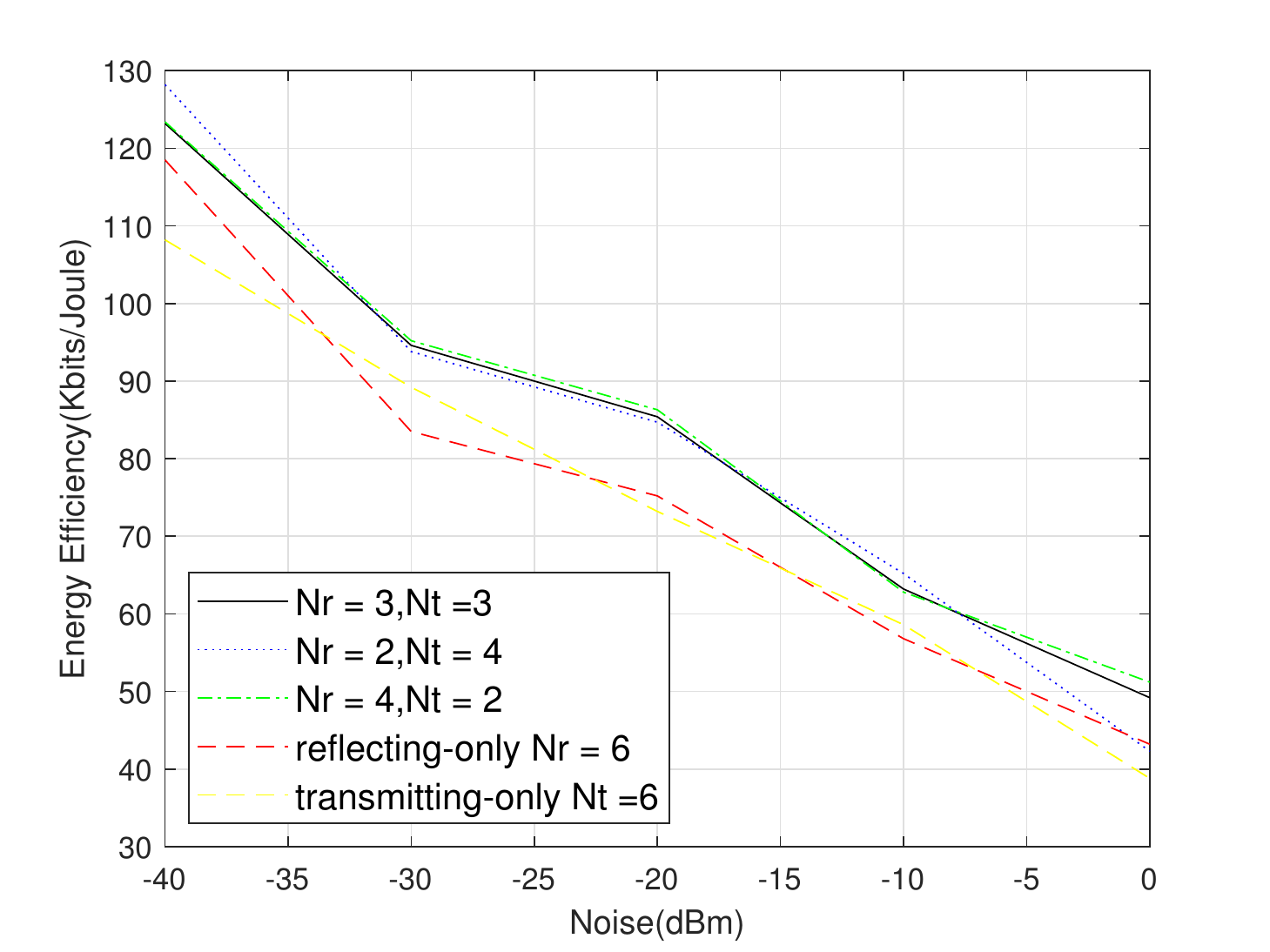}\vspace{-2mm}
		\caption{The minum user EE versus system noise.}
		\label{fig:Estimation_Scheme} \vspace{-0.5mm}
	\end{minipage}
	\begin{minipage}{9cm}
		\centering
		\includegraphics[width=8cm]{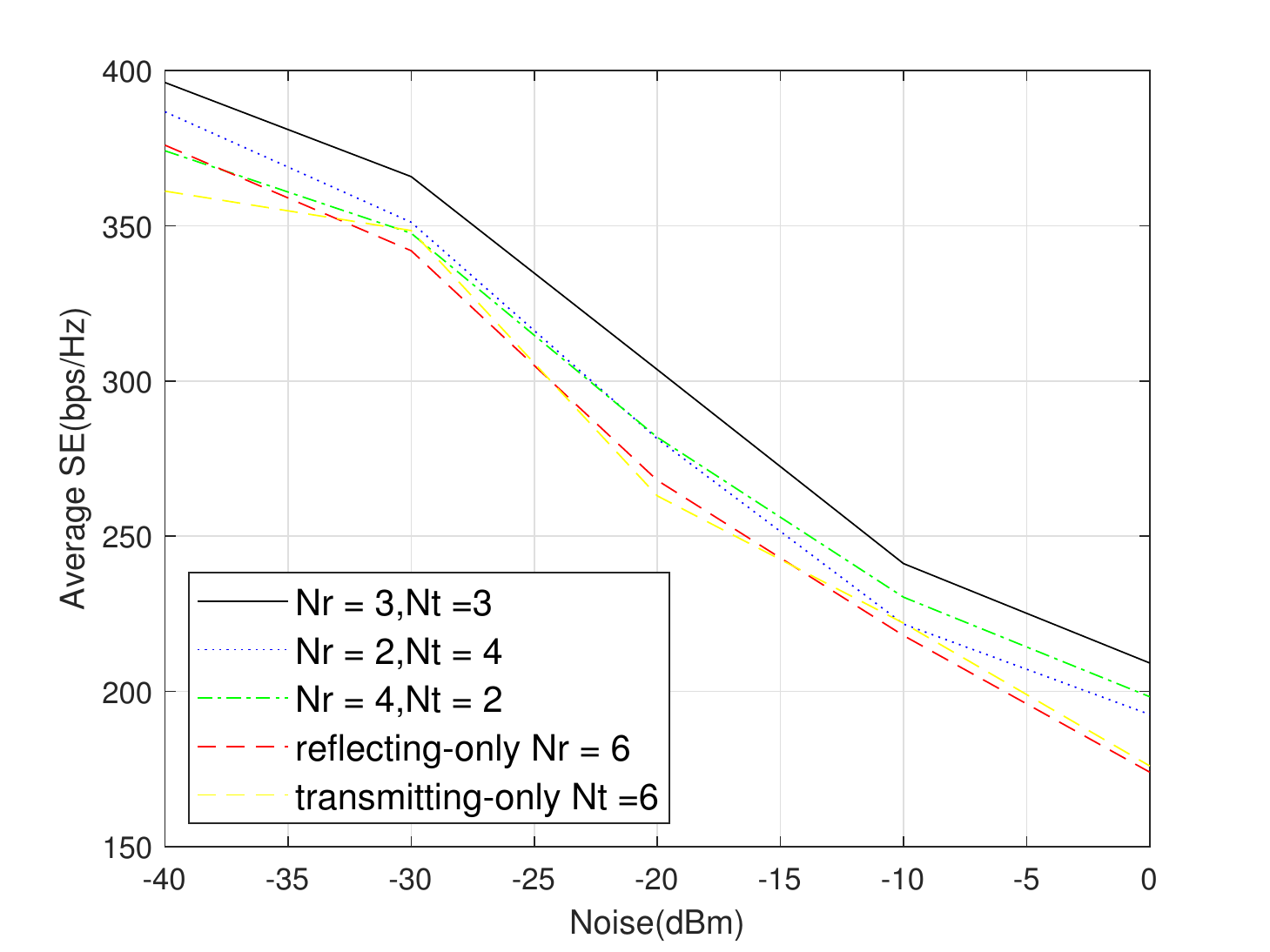}\vspace{-2mm}
		\caption{The average SNR system noise.}
		\label{fig:Estimation_Scheme}\vspace{-1.5mm}
	\end{minipage}
\end{figure}
Fig 6 and 7 show the changes of EE and average SE when the system noise increases respectively. The increase of noise in communication system seriously affects the signal transmission quality, reduces the SNR ratio, and thus lower the user EE. It is not difficult to find that STAR-RIS assisted system performs better when facing larger noise power.

\bibliographystyle{IEEEtran}
\bibliography{reference}
\end{document}